\documentclass[12pt]{article}
\usepackage[margin=1in]{geometry}
\usepackage[T1]{fontenc}
\usepackage[utf8]{inputenc}
\usepackage{lmodern}
\usepackage{microtype}
\usepackage{amsmath,amssymb}
\usepackage{booktabs}
\usepackage{array}
\usepackage{graphicx}
\usepackage[font=small,labelfont=bf]{caption}
\usepackage{placeins}
\usepackage{xurl}
\usepackage{hyperref}
\usepackage{float}
\hypersetup{
  colorlinks=true,
  linkcolor=black,
  citecolor=black,
  urlcolor=blue,
  pdftitle={A Constraint-Modulated Rate Law for Glass-Forming Systems: Viscosity Evidence and Dielectric Transfer in Salol},
  pdfauthor={Debra S. Gavant and Christian E. Precker}
}

\graphicspath{{figures/}}

\title{\vspace*{-1.1 em}\textbf{A Constraint-Modulated Rate Law for Glass-Forming Systems:\\ \Large Viscosity Evidence and Dielectric Transfer in Salol}}

\author{
Debra S. Gavant\thanks{Corresponding author: \href{mailto:dsgavant@gmail.com}{dsgavant@gmail.com}}\\
\small DP$\Phi$ Research Initiative, Atlanta, Georgia, USA
\and
Christian E. Precker\thanks{\href{mailto:christian.precker@aimen.es}{christian.precker@aimen.es}}\\
\small AIMEN Technology Centre, O Porri\~no, Spain
}
\date{\small August 22, 2026}

\begin{document}
\maketitle
\renewcommand{\arraystretch}{1.18}

\vspace*{-.5 em}
\begin{abstract}
The viscosity of a glass-forming liquid can increase by more than 14 decades as structural mobility diminishes upon cooling. Continuous Present Actualization (CPA) treats physical change as an ongoing process under the physical conditions currently in force. CPA plus constraint (CPA+C) tests the material hypothesis that relational constraints modulate the observable rate of structural reconfiguration. The bounded function $C(T)$ supplies an operational temperature modulation for this proposed influence. The five-parameter law is compared with the Vogel--Fulcher--Tammann (VFT), Mauro--Yue--Ellison--Gupta--Allan (MYEGA), and Avramov--Milchev equations using source-verified data. On a salol viscosity record comprising 35 published data symbols and spanning 14.15 decades, CPA+C is favored over the nearest comparator by 47.72 Akaike information criterion (AIC) units and reduces the leave-one-symbol-out root-mean-square prediction error from 0.398 to 0.207. It is selected in all 1,000 symbol-level bootstrap resamples under AIC, its small-sample correction (AICc), and the Bayesian information criterion (BIC). The same construction is decisively selected on 60 source-verified boron trioxide (B$_2$O$_3$) viscosity measurements, with a 57.39-unit AIC advantage over the nearest comparator. Salol then provides the prospectively identified cross-observable test. The viscosity fit fixes both temperature coordinates and the complete temperature dependence of the viscosity-derived modulation before the dielectric comparison; only dielectric-specific coefficients are fitted. This fixed-geometry model improves on a denominator-matched control by 16.32 AIC units, and the AIC preference persists across all 1,000 propagated viscosity geometries. The viscosity-derived modulation therefore retains predictive information for molecular reorientation in the same liquid. Together, the salol and B$_2$O$_3$ viscosity results and the salol transfer establish a reproducible constraint-modulated rate signature consistent with progressive narrowing of configurational access.
\end{abstract}

\section{Introduction}

Glass-forming liquids can increase their viscosity by more than 14 decades over a limited temperature range. This rise reflects a substantial reduction in structural mobility. Several compact models reproduce the resulting curvature with high empirical accuracy. The Vogel--Fulcher--Tammann (VFT) equation remains the historical reference~\cite{Vogel1921,Fulcher1925,Tammann1926}. The Mauro--Yue--Ellison--Gupta--Allan (MYEGA) equation provides an energy-landscape and constraint-counting formulation without a finite-temperature divergence~\cite{Mauro2009}. Avramov--Milchev offers an entropy-based alternative with the same three-parameter complexity~\cite{Avramov1988}. A useful alternative must therefore do more than add flexibility: it should connect the curvature to a testable physical process and retain its empirical advantage under complexity penalties, held-out prediction, resampling, and transfer to a second observable.

CPA plus constraint (CPA+C) is an operational rate law motivated by \textit{Continuous Present Actualization} (CPA) within \textit{Dynamic Present Theory} (DP$\Phi$)~\cite{GavantDPTI}. In this framework, the liquid reorganizes under its current physical conditions, which restrict the molecular rearrangements available at a given temperature. CPA+C tests the material consequence of this proposal: cooling-induced narrowing of those rearrangements should leave a bounded, temperature-dependent signature in structural slowing. The modulation $C(T)$ supplies the operational temperature dependence tested here.

Viscosity measures the liquid's macroscopic resistance to flow, whereas dielectric $\alpha$-relaxation tracks molecular dipole reorientation. These observables probe different motions and can couple differently to the liquid's structural slowing. Density-scaling studies have shown that reduced viscosity and reorientational relaxation times can be superposed on a common master curve when plotted against the combined temperature-density variable $T/\rho^\gamma$ in salol and other non-associated liquids~\cite{Casalini2016}. Earlier work on salol also compared dielectric relaxation directly with viscosity, including the relation $\tau_D \propto \eta/T$~\cite{Hansen1998}. The more specific question addressed here is whether a bounded temperature modulation inferred from viscosity retains predictive value for dielectric relaxation when its geometry is fixed before the dielectric fit. This cross-observable test was identified prospectively in DP$\Phi$ I v6~\cite{GavantDPTIv6Prospective}.

The empirical program proceeds in two stages. A salol viscosity record comprising 35 published data symbols and spanning 14.15 decades supports penalized comparison, symbol-level resampling, and leave-one-symbol-out prediction~\cite{Laughlin1972}. The boron trioxide (B$_2$O$_3$) record comprises 60 source-verified viscosity measurements across three viscometer configurations and extends the comparison from a fragile molecular liquid to a covalent network glass former~\cite{Napolitano1965}. Salol then supplies the within-material transfer test. Its viscosity fit fixes both temperature coordinates and the complete temperature dependence of the modulation before the dielectric comparison; only dielectric-specific coefficients are estimated. A matched control retains the viscosity-fixed denominator but omits $C(T)$. The comparison therefore tests whether the viscosity-derived modulation adds predictive information for dielectric relaxation, rather than merely whether CPA+C can fit both observables independently.

\section{Constraint-Modulated Rate Law}

\subsection{Physical motivation}

As a glass-forming liquid cools, changes in packing and local coordination, together with geometric frustration, narrow the range of molecular arrangements it can access. As independent molecular motions become more restricted, structural change increasingly requires cooperative rearrangements. In network glass formers, rigidity imposes an additional constraint on mobility. Adam--Gibbs theory relates slowing to configurational entropy and cooperatively rearranging regions~\cite{Adam1965}; Phillips--Thorpe theory describes rigidity through mechanical constraint counting~\cite{Phillips1979,Thorpe1983}; mode-coupling theory identifies a regime of sharply slowing relaxation above the laboratory glass transition~\cite{Gotze1992}; and energy-landscape descriptions organize relaxation around accessible states and barriers~\cite{Mauro2009}. These established accounts identify the material conditions under which structural relaxation slows.

CPA interprets these material conditions as part of the physical process of reorganization. Molecular rearrangements proceed through motions compatible with those conditions, and cooling narrows that compatible set. CPA+C represents the proposed rate effect through the bounded modulation $C(T)$.

The model is evaluated through three progressively more demanding tests: penalized comparison with established rate equations, prediction of held-out viscosity observations, and fixed-geometry transfer to a second observable against a control with the same denominator. The first two determine whether the curvature introduced by $C(T)$ is reproducible in viscosity. The third asks whether the temperature dependence inferred from viscosity remains predictive for molecular reorientation in the same liquid.

\subsection{Operational form}

For viscosity, the CPA+C law is

\begin{equation}
\log_{10}\eta(T)
=
a+\frac{A+B\,C(T)}{T-T_{\mathrm{lock}}},
\label{eq:cpac}
\end{equation}

\vspace{.5em}
where $a$ is a logarithmic prefactor, $A$ is the baseline kinetic amplitude, and $B$ is the constraint-coupling amplitude. The two fitted temperature coordinates are $T_{\mathrm{lock}}$ and $T_{\mathrm{ref}}$. The bounded modulation is

\begin{equation}
C(T)
=
\mathrm{clip}\!\left(
\frac{T_{\mathrm{ref}}-T_{\mathrm{lock}}}
{T-T_{\mathrm{lock}}},
0,1
\right),
\label{eq:constraint}
\end{equation}

where $\mathrm{clip}(x,0,1)$ leaves values between zero and one unchanged and sets values outside that interval to the nearest bound. Together, $a$, $A$, $B$, $T_{\mathrm{lock}}$, and $T_{\mathrm{ref}}$ give the model five fitted parameters.

Equation~\eqref{eq:cpac} separates a baseline term from a constraint-coupled term:

\[
\log_{10}\eta(T)
=
a
+
\frac{A}{T-T_{\mathrm{lock}}}
+
\frac{B\,C(T)}{T-T_{\mathrm{lock}}}.
\]

\vspace{.5em}
The first fraction supplies a VFT-like baseline dependence. The second contributes the additional temperature dependence associated with the proposed narrowing of configurational access. At each temperature, $C(T)$ determines the fraction of the fitted amplitude $B$ entering the numerator, while $B$ sets the strength of that contribution. For viscosity, Eq.~\eqref{eq:cpac} predicts $\log_{10}\eta$; in the transfer analysis, the same construction predicts $\log_{10}\tau_\alpha$ for dielectric relaxation. Larger values correspond to slower structural or molecular dynamics.

Within DP$\Phi$, molecular reorganization depends on the liquid’s physical organization as a whole. The rate law represents the temperature dependence of that proposed effect through $C(T)$, which serves as a scalar operational coordinate for the empirical test.

Over the fitted domain $T>T_{\mathrm{lock}}$, substituting Eq.~\eqref{eq:constraint} into Eq.~\eqref{eq:cpac} makes the source of the additional curvature explicit:

\[
\log_{10}\eta(T)
=
\begin{cases}
\displaystyle
a+\frac{A+B}{T-T_{\mathrm{lock}}},
& T_{\mathrm{lock}}<T\leq T_{\mathrm{ref}},
\\[2ex]
\displaystyle
a+\frac{A}{T-T_{\mathrm{lock}}}
+\frac{B\left(T_{\mathrm{ref}}-T_{\mathrm{lock}}\right)}
{\left(T-T_{\mathrm{lock}}\right)^2},
& T>T_{\mathrm{ref}}.
\end{cases}
\]

\vspace{.5em}
For $T>T_{\mathrm{ref}}$, cooling increases $C(T)$ toward unity. The constraint-coupled term then has an inverse-square temperature dependence, adding curvature beyond the baseline term. At $T_{\mathrm{ref}}$, the modulation reaches one. Within the fitted domain $T_{\mathrm{lock}}<T\leq T_{\mathrm{ref}}$, the full amplitude $B$ contributes, and the law takes the VFT form with numerator $A+B$. As temperature decreases toward $T_{\mathrm{lock}}$, the denominator $T-T_{\mathrm{lock}}$ becomes smaller, so $\log_{10}\eta(T)$ continues to rise even though $C(T)$ has saturated at unity.

This change in curvature is the distinctive empirical feature introduced by CPA+C. VFT, MYEGA, and Avramov--Milchev each impose their own curvature across the fitted range. CPA+C allows the constraint-coupled term to vary above $T_{\mathrm{ref}}$, saturate at $T_{\mathrm{ref}}$, and leave continued slowing below that boundary to the VFT-like denominator. The model comparison therefore tests whether the data contain this specific curvature transition strongly enough to justify the two additional parameters.

The two temperature coordinates have distinct roles. $T_{\mathrm{ref}}$ locates the boundary at which the modulation saturates, while $T_{\mathrm{lock}}$ sets the denominator scale. Together with the resulting shape of $C(T)$, they define the model's temperature geometry. Both are fitted coordinates inferred from the measured rate trajectory.

\begin{figure}[htbp]
\centering
\includegraphics[width=\linewidth]{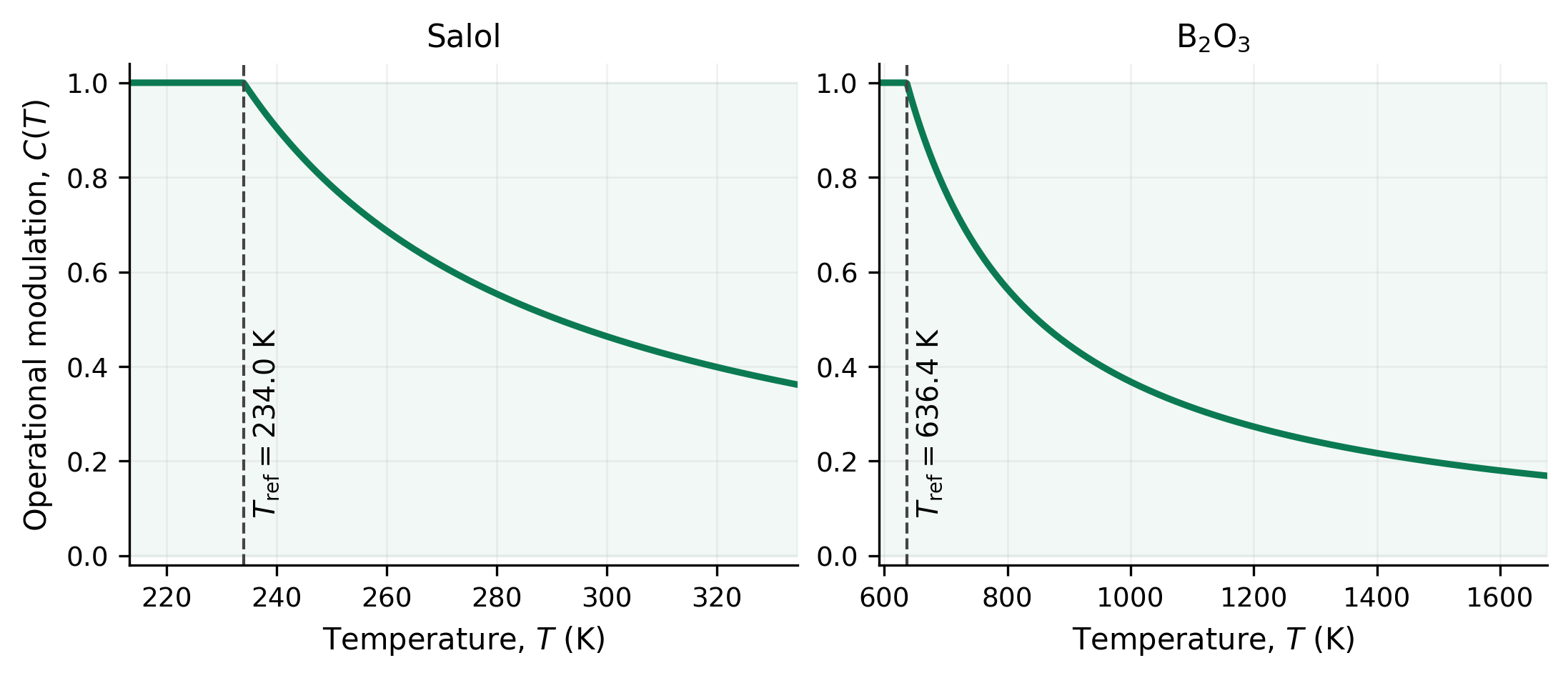}
\caption{Operational modulation $C(T)$ inferred separately for salol and B$_2$O$_3$. Dashed lines mark the fitted $T_{\mathrm{ref}}$ values. Within the plotted domain, $C(T)=1$ for $T_{\mathrm{lock}}<T\leq T_{\mathrm{ref}}$; for $T>T_{\mathrm{ref}}$, the modulation decreases as temperature rises. The same bounded form is shown on each material's independently fitted temperature scale.}
\label{fig:constraint}
\end{figure}

Figure~\ref{fig:constraint} shows the fitted modulation for salol and B$_2$O$_3$. Both materials use the same bounded form, with temperature coordinates fitted independently for each material. For salol, $T_{\mathrm{lock}}$, $T_{\mathrm{ref}}$, and the resulting $C_\eta(T)$ form the fixed temperature geometry of the dielectric transfer; the dielectric coefficients are fitted separately.

\section{Analysis Design}

\subsection{Data roles and source resolution} 

The viscosity records of salol and B$_2$O$_3$ serve as the primary datasets for model comparison. Salol also supplies the data for the direct dielectric comparison and the within-material, fixed-geometry transfer test. Table~\ref{tab:data} summarizes the three records included in the quantitative analysis.

\begin{table}[!ht] 
\centering 
\caption{Primary records used in the quantitative analysis.} \label{tab:data} \small \setlength{\tabcolsep}{3.5pt} \begin{tabular}{>{\raggedright\arraybackslash}p{0.21\linewidth}>{\raggedright\arraybackslash}p{0.15\linewidth}>{\raggedright\arraybackslash}p{0.22\linewidth}>{\raggedright\arraybackslash}p{0.24\linewidth}} 
\toprule Record & $n$ & Range & Role \\ \midrule Salol viscosity, Laughlin--Uhlmann Fig.~3 & 35 symbols & 213.31--334.67 K; 14.15 decades & Model comparison, bootstrap, leave-one-symbol-out prediction, and viscosity-derived temperature geometry for transfer \\ B$_2$O$_3$, Napolitano Table~I & 60 measurements & 591.45--1676.65 K; 8.83 decades & Network-glass comparison across three viscometer configurations \\ Salol dielectric, Lunkenheimer et al. & 21 equilibrium observations; 10-observation transfer subset & 223--361 K equilibrium; 223--265 K transfer & Dielectric-only model comparison; viscosity-to-dielectric transfer with viscosity-derived temperature geometry held fixed \\ \bottomrule \end{tabular} 
\end{table} 

Before the combined analysis, all viscosity inputs were re-examined against their primary publications. The audit found that the available ortho-terphenyl records have overlapping source histories and differ in how they were assembled. They are preserved in the accompanying archive for provenance and source-construction review.

The salol viscosity analysis uses the 35 symbols published in Figure~3 of Laughlin and Uhlmann~\cite{Laughlin1972}. Each symbol enters the fit once and serves as the unit for bootstrap resampling and leave-one-symbol-out prediction. An independent reconstruction of the same record supported the validation workflow; earlier coordinate-level reconstructions are retained in the archive for historical provenance.

The B$_2$O$_3$ input is a direct transcription of the 60 viscosity measurements reported in Table~I of Napolitano, Macedo, and Hawkins~\cite{Napolitano1965}: nine driving-inner, 11 decay-inner, and 40 driving-outer measurements. Table~II is not used because its viscosity values were graphically interpolated rather than reported as a separate observational series. 

The salol dielectric analysis uses 21 equilibrium observations spanning 223--361~K for the direct model comparison and the 10-observation subset spanning 223--265~K for the transfer test~\cite{Lunkenheimer2010}. The 212~K observation belongs to a separately defined aging regime. The transfer window contains three observations below and seven above the viscosity-derived $T_{\mathrm{ref}}$. Among the records assembled for this study, salol alone provided equilibrium viscosity and dielectric $\alpha$-relaxation data with sufficient range and overlap for the fixed-geometry transfer design.

\vspace{-2mm}
\subsection{Comparator models and fitting protocol} 

The three comparator laws are
\begin{align} 
\text{VFT:}\quad & y(T)=a+\frac{B}{T-T_0}, \\ 
\text{MYEGA:}\quad & y(T)=a+\frac{K}{T}\exp\!\left(\frac{C}{T}\right), \\ 
\text{Avramov--Milchev:}\quad & y(T)=a+\left(\frac{\Omega}{T}\right)^{\alpha},
\end{align} 

where $y=\log_{10}\eta$ for viscosity or $y=\log_{10}\tau_\alpha$ for dielectric relaxation. Each comparator has three fitted parameters. The parameter symbols are model-specific; in particular, the MYEGA parameter $C$ is distinct from the CPA+C modulation $C(T)$. A constant multiplicative factor in the Avramov--Milchev term can be absorbed into $\Omega$ without changing the model family. 

All models are fitted by minimizing the equally weighted residual sum of squares in $y$. Primary point fits use bounded nonlinear least squares with 300 reproducible randomized starting points per model. All free logarithmic prefactors use the common primary bound $a\in[-20,5]$. Complete model-specific bounds, starting-point conventions, numerical tolerances, software versions, and execution records are preserved in the accompanying archive. CPA+C retains its nominal $k=5$ parameter count in every information-criterion calculation, including when a fitted parameter lies at an imposed bound. 

The criteria used for model comparison are the Akaike information criterion (AIC), its small-sample correction (AICc), and the Bayesian information criterion (BIC): 
\vspace{.5em}
\begin{align} 
\mathrm{AIC}&=n\ln(\mathrm{RSS}/n)+2k,\\ \mathrm{AICc}&=\mathrm{AIC}+\frac{2k(k+1)}{n-k-1},\\ \mathrm{BIC}&=n\ln(\mathrm{RSS}/n)+k\ln n, 
\end{align} 

where $\mathrm{RSS}$ is the residual sum of squares, $n$ is the number of observations, and $k$ is the fitted parameter count. For each record and criterion, the reported $\Delta$ is the model's score minus the minimum across the four models. Positive $\Delta$ values indicate less support than the best-supported model under that criterion. A $\Delta\mathrm{AIC}>10$ conventionally indicates strong support for the lowest-AIC model~\cite{Burnham2002}.

The initial viscosity test against VFT and its success criterion were preregistered and archived before fitting in \emph{Glass-Freeze Analysis Protocol v0}~\cite{GavantGlassProtocol}. MYEGA and Avramov--Milchev were subsequently added as established three-parameter comparators, extending the analysis beyond VFT. The same four-model protocol is applied to each record entering a direct model comparison.

\vspace{-2mm}
\subsection{Salol resampling and held-out prediction} 

The primary salol bootstrap generates 1,000 symbol-level resamples. Each resample contains 35 draws with replacement from the 35 salol symbols; individual symbols may recur while others are omitted. All four models are fitted to every resample using 50 randomized starting points per fit. The primary analysis uses the prefactor bound $a\in[-20,5]$. Because fitted prefactors crossed the predeclared trigger, the specified secondary analysis repeats the 1,000-resample comparison under the alternative prefactor bound $a\in[-10,5]$.

The reported 95\% parameter intervals are calculated from percentiles of the fitted symbol-level resamples and quantify sampling variability among the 35 salol symbols. Experimental measurement error and manual digitization error were not assigned probability models in the specified workflow and are therefore not represented in these intervals.

Predictive performance is evaluated by leave-one-symbol-out cross-validation. Each of the 35 symbols is withheld once; all four models are refitted to the remaining 34 symbols, and the held-out logarithmic viscosity is predicted. These parameter counts, bounds, and fitting conventions are retained across all 35 folds.

\vspace{-3mm}
\subsection{Salol transfer with fixed viscosity geometry} 

The salol viscosity point fit determines 
$T_{\mathrm{lock},\eta}$, $T_{\mathrm{ref},\eta}$, and the complete temperature dependence of $C_\eta(T)$. Let 
\[ 
R_\eta(T)= 
\frac{A_\eta+B_\eta C_\eta(T)} 
{T-T_{\mathrm{lock},\eta}} 
\] 
denote the temperature-dependent part of the fitted viscosity law, and let $y_\epsilon(T)=\log_{10}\tau_\alpha(T)$ denote the dielectric observable. The transfer forms are 
\begin{align} 
\mathrm{Q2a:}\quad y_\epsilon(T) &= a_\epsilon+s_\epsilon R_\eta(T),\\ 
\mathrm{Q2b:}\quad y_\epsilon(T) &= a_\epsilon+\frac{A_\epsilon+B_\epsilon C_\eta(T)}{T-T_{\mathrm{lock},\eta}},\\ \mathrm{CTRL:}\quad y_\epsilon(T) &= a+\frac{b}{T-T_{\mathrm{lock},\eta}}. 
\end{align} 
The transfer analysis includes a strict affine diagnostic (Q2a), a fixed-geometry model with dielectric-specific coupling (Q2b), and a denominator-matched control (CTRL). Q2a applies one offset and one scale factor to the temperature-dependent part of the fitted viscosity law. Q2b holds $T_{\mathrm{lock},\eta}$, $T_{\mathrm{ref},\eta}$, and $C_\eta(T)$ fixed while fitting three dielectric-specific coefficients: $a_\epsilon$, $A_\epsilon$, and $B_\epsilon$. CTRL retains the same viscosity-fixed denominator as Q2b, $T-T_{\mathrm{lock},\eta}$, but omits $C_\eta(T)$; its CTRL-specific coefficients $a$ and $b$ are fitted to the dielectric data. The primary transfer comparison between Q2b and CTRL tests whether the viscosity-derived modulation provides predictive information beyond the viscosity-fixed denominator alone.

Q2a and CTRL each have two dielectric-fitted parameters; Q2b has three. The primary AICc calculation counts those dielectric-fitted parameters. A sensitivity calculation also counts the residual-variance parameter, allowing the dependence of the Q2b--CTRL comparison on the parameter-count convention to be assessed. 

Because CTRL has one fewer dielectric-fitted parameter than Q2b, the information criteria favor Q2b only when its improved fit offsets the penalty for the additional coupling amplitude.

All 1,000 viscosity geometries generated by the primary symbol-level bootstrap are propagated through the transfer analysis. No geometry is modified or discarded. The final 35-symbol workflow was independently reproduced, including the point fits, primary and secondary bootstrap analyses, and all 1,000 transfer evaluations, with no numerical mismatches or failed evaluations. The signed review records the verdict \emph{Validated with Qualifications}; the full report and addendum are preserved in the accompanying archive. For the final execution, the 35-symbol input, symbol-level resampling rule, model equations, dielectric window, denominator-matched control definition, parameter bounds, starting-point counts, and random seeds were fixed and hash-recorded on August~3, 2026, before the analysis was conducted.

\section{Results}

\subsection{Salol viscosity} 

\begin{figure}[H] 
\centering 
\includegraphics[width=1.\linewidth]{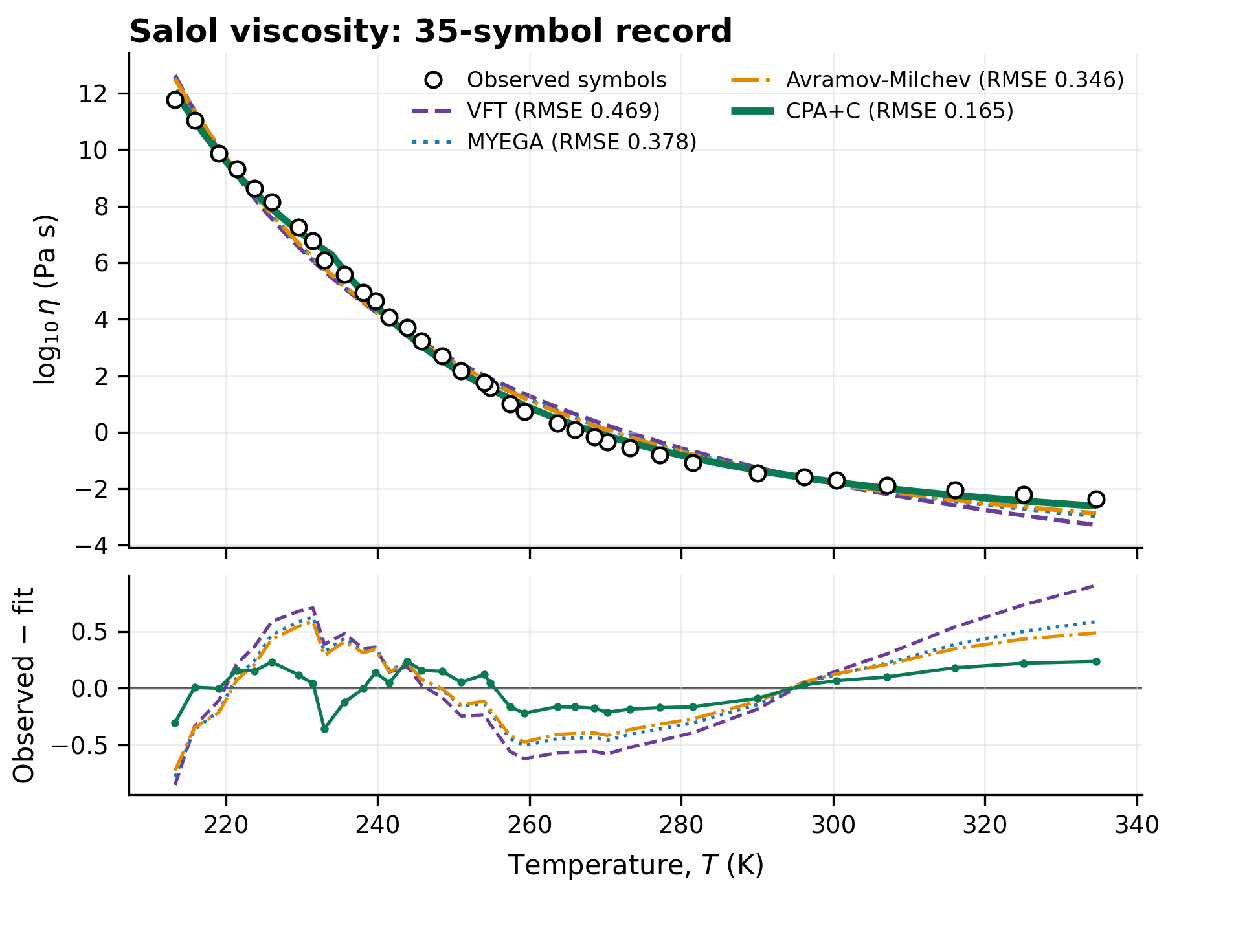} 
\vspace{-30pt}
\caption{Four-model comparison using the salol viscosity symbols digitized from Figure~3 of Laughlin and Uhlmann~\cite{Laughlin1972}. The upper panel shows fitted curves; the lower panel shows observed-minus-fit residuals. Viscosity is expressed in Pa\,s.} \label{fig:salol} 
\end{figure}

Across the 14.15-decade salol trajectory, CPA+C leaves substantially less systematic residual curvature than the three comparator equations (Fig.~\ref{fig:salol}). Its RMSE is 0.1651, compared with 0.3457 for the nearest comparator, Avramov--Milchev. With the nominal five-parameter penalty retained, CPA+C is favored over Avramov--Milchev by 47.724 AIC units, 46.429 AICc units, and 44.613 BIC units; the margins over MYEGA and VFT are larger still (Table~\ref{tab:mainfit}).

The CPA+C point fit gives 
\begin{equation} 
T_{\mathrm{lock}}=176.9716\ \mathrm{K},\qquad T_{\mathrm{ref}}=234.0078\ \mathrm{K}. 
\end{equation} 

The corresponding symbol-bootstrap 95\% intervals are 
\begin{align} 
T_{\mathrm{lock}}&:\ [170.4699,184.9806]\ \mathrm{K},\\ T_{\mathrm{ref}}&:\ [230.3758,237.8023]\ \mathrm{K}. 
\end{align} 

\begin{table}[!ht] 
\centering 
\caption{Penalized point-fit comparison for the principal viscosity records. RMSE values are reported on the $\log_{10}\eta$ scale. All $\Delta$ values are relative to CPA+C within the same record.}
 \vspace{1em}
\label{tab:mainfit} 
\small \begin{tabular}{llrrrr} 
\toprule 
Record & Model & RMSE & $\Delta$AIC & $\Delta$AICc & $\Delta$BIC \\ \midrule 
Salol ($n=35$) & VFT & 0.4690 & 69.074 & 67.779 & 65.963 \\ 
& MYEGA & 0.3776 & 53.900 & 52.605 & 50.789 
\\ & Avramov--Milchev & 0.3457 & 47.724 & 46.429 & 44.613 
\\ & CPA+C & \textbf{0.1651} & \textbf{0} & \textbf{0} & \textbf{0} \\ 
\addlinespace 
B$_2$O$_3$ ($n=60$) & VFT & 0.0628 & 57.389 & 56.707 & 53.201 \\ 
& MYEGA & 0.0663 & 63.901 & 63.219 & 59.713 \\ 
& Avramov--Milchev & 0.0979 & 110.730 & 110.048 & 106.542 \\ 
& CPA+C & \textbf{0.0376} & \textbf{0} & \textbf{0} & \textbf{0} \\ \bottomrule 
\end{tabular} 
\end{table} 

\vspace{2em}

In the point fit, the baseline amplitude $A=5.91\times10^{-29}$ lies at its zero lower bound; the same bound is reached in 953 of the 1,000 primary resamples. All information-criterion calculations retain $k=5$, so the reported model-selection advantage incurs the full nominal complexity penalty.

\begin{figure}[!ht] 
\centering 
\includegraphics[width=\linewidth]{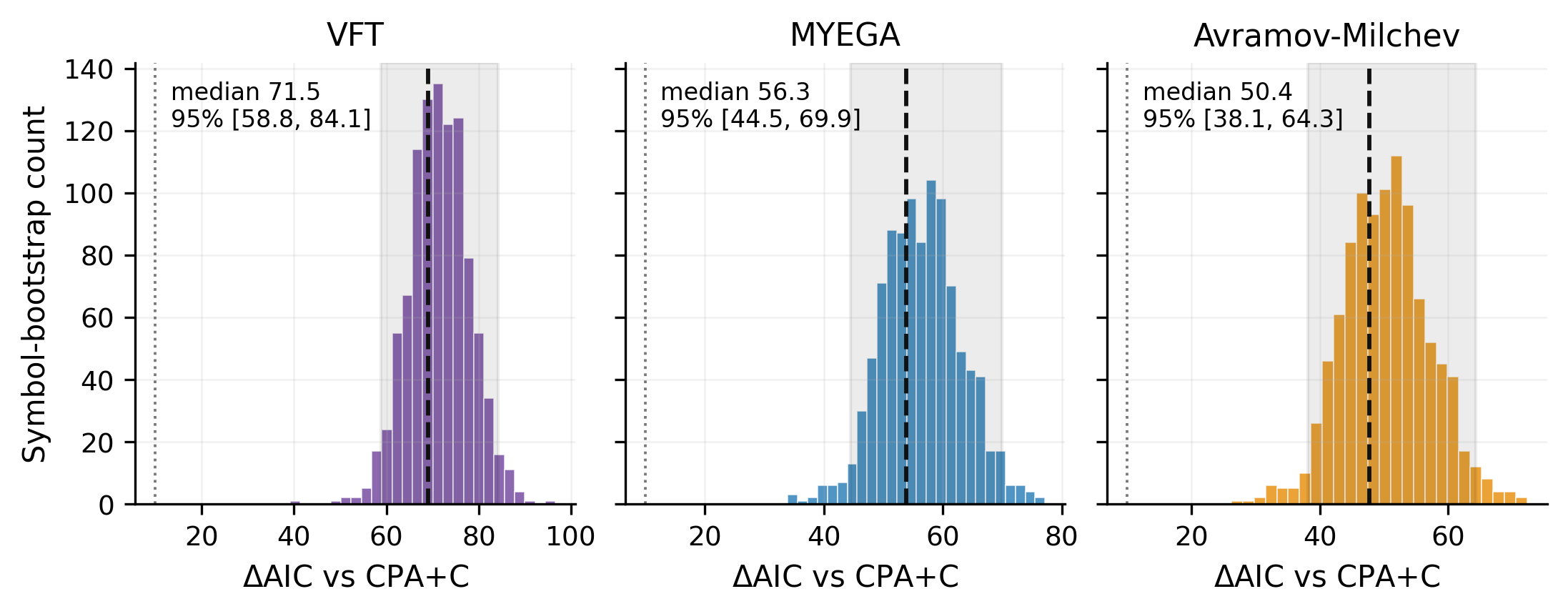} \caption{Symbol-bootstrap $\Delta$AIC distributions for the salol comparators relative to CPA+C. Shading marks the central 95\% interval, dashed black lines mark point-fit $\Delta$AIC values, and dotted gray lines mark $\Delta\mathrm{AIC}=10$.} \label{fig:bootstrap} \end{figure}

CPA+C is selected by AIC, AICc, and BIC in all 1,000 primary resamples and all 1,000 secondary-bound resamples, with no failed fits. For the primary bootstrap, the median $\Delta\mathrm{AIC}$ values relative to CPA+C, with 95\% intervals in brackets, are 71.50 $[58.83,84.14]$ for VFT, 56.26 $[44.49,69.90]$ for MYEGA, and 50.38 $[38.14,64.30]$ for Avramov--Milchev (Fig.~\ref{fig:bootstrap}). In every primary resample, each comparator remains more than 10 AIC units above CPA+C. 

\vspace{2em}
\begin{table}[!ht] 
\centering 
\caption{Leave-one-symbol-out prediction errors for salol. Errors are in $\log_{10}\eta$.} 
\label{tab:loo} 
\small \begin{tabular}{lrr} 
\toprule 
Model & MAE & RMSE \\ 
\midrule 
VFT & 0.4603 & 0.5460 \\ 
MYEGA & 0.3715 & 0.4367 \\ 
Avramov--Milchev & 0.3396 & 0.3982 \\ 
CPA+C & \textbf{0.1701} & \textbf{0.2065} \\
\bottomrule 
\end{tabular} 
\end{table} 

\vspace{1em}
All four models completed every held-out fold. CPA+C has the lowest leave-one-symbol-out MAE and RMSE (Table~\ref{tab:loo}); its RMSE of 0.2065 is approximately half the nearest comparator RMSE of 0.3982.

\FloatBarrier

\vspace*{2mm}

\subsection{Boron trioxide viscosity}

\begin{figure}[!ht]
\centering
\includegraphics[width=.96\linewidth]{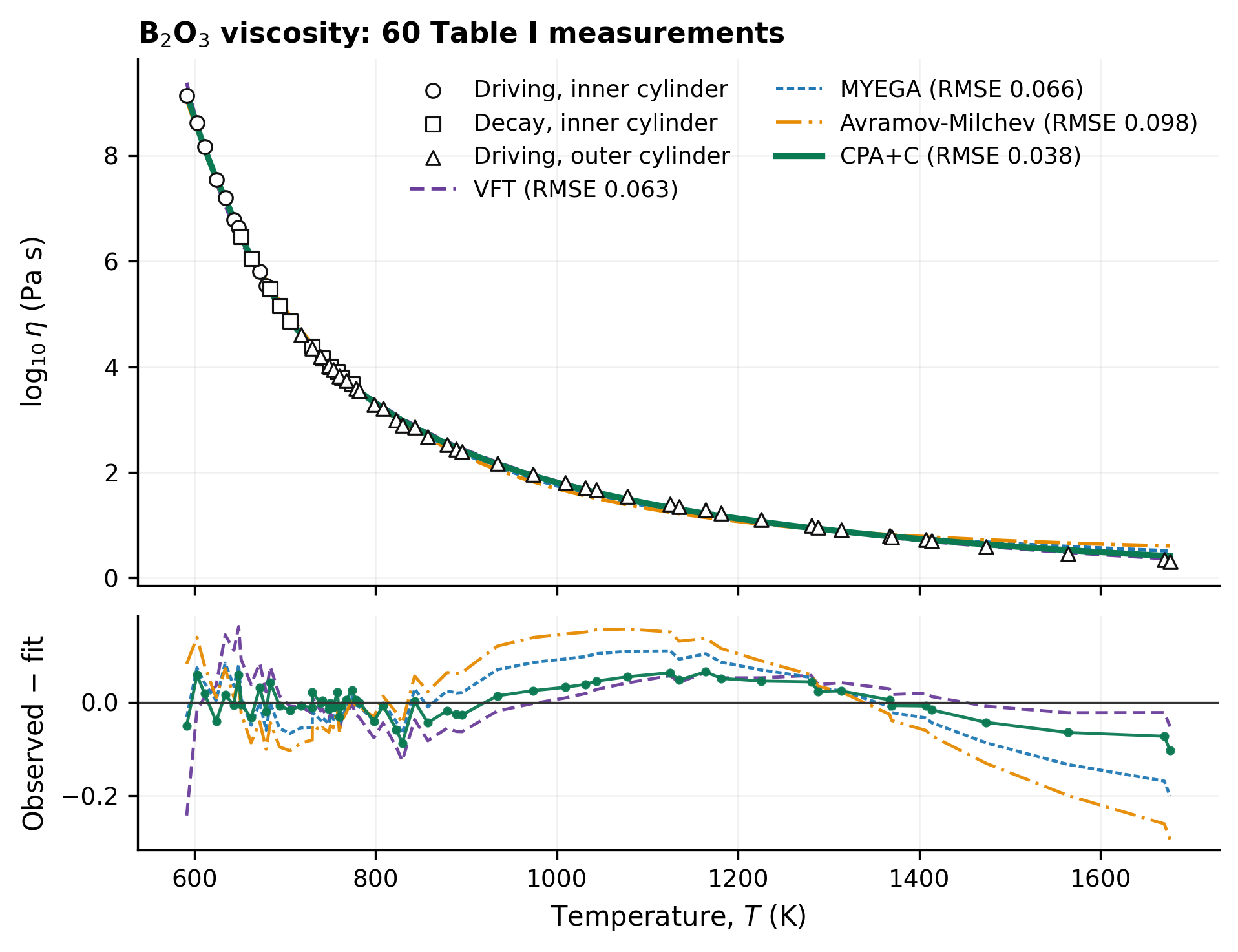}
\caption{Four-model comparison for the source-verified B$_2$O$_3$ Table~I record. Marker shapes identify the three reported viscometer configurations. The upper panel shows fitted curves; the lower panel shows observed-minus-fit residuals.}
\label{fig:b2o3}
\end{figure}

The source-verified B$_2$O$_3$ record extends the comparison from a fragile molecular liquid to a covalent network glass former. CPA+C gives a point-fit RMSE of 0.03763 and is favored over VFT by 57.389 AIC units, over MYEGA by 63.901, and over Avramov--Milchev by 110.730. AICc and BIC give the same ranking (Table~\ref{tab:mainfit}). Residuals remain small across the three reported viscometer configurations (Fig.~\ref{fig:b2o3}).

\vspace{1em}
The CPA+C point fit gives
\[
T_{\mathrm{lock}}=424.7730\ \mathrm{K},\qquad
T_{\mathrm{ref}}=636.3627\ \mathrm{K}.
\]
Both fitted amplitudes lie well within their allowed bounds:
\[
A=1295.8163\ \mathrm{K},\qquad
B=346.5765\ \mathrm{K},
\]
so both the baseline and modulation terms contribute. Together, $T_{\mathrm{lock}}$ and $T_{\mathrm{ref}}$ locate the denominator scale and the modulation-saturation boundary in the fitted B$_2$O$_3$ law.

\FloatBarrier
\subsection{Salol dielectric transfer and comparison}

The fixed-geometry transfer comparison provides the primary dielectric result. The salol viscosity fit fixes
$T_{\mathrm{lock},\eta}=176.9716$~K,
$T_{\mathrm{ref},\eta}=234.0078$~K,
and the complete temperature dependence of $C_\eta(T)$. In the 10-observation transfer window, Q2b gives a point-fit RMSE of $0.03605$, compared with $0.09010$ for the denominator-matched control. For reference, a dielectric-only CPA+C benchmark that refits all CPA+C parameters on the same 10-observation window gives an RMSE of $0.0297$. Relative to CTRL, Q2b is favored by 16.3214 AIC units and 16.0189 BIC units. The corresponding AICc advantage is 14.0357 units under the primary parameter-count convention and 12.3214 units when the residual-variance parameter is also counted.

Across the 1,000 propagated viscosity geometries, Q2b is favored over CTRL in all 1,000 cases by AIC, BIC, and the primary AICc calculation, and in 998 cases under the alternative AICc calculation. The propagated $\Delta\mathrm{AIC}$ distribution has a median of 14.3869 and a central 95\% interval of $[7.4760,25.9233]$; 81.1\% of propagated margins exceed 10.

\begin{figure}[!ht]
\centering
\includegraphics[width=.89\linewidth]{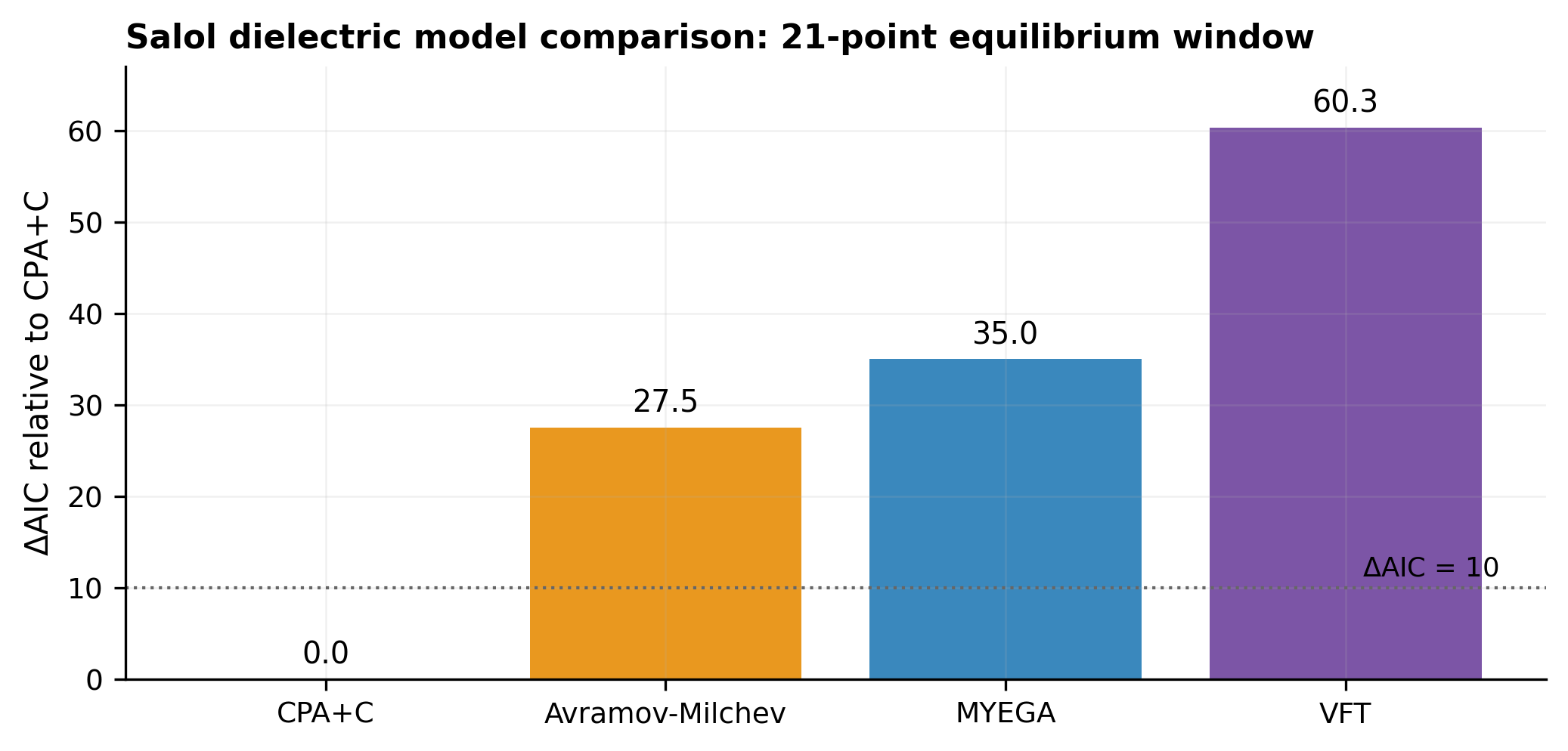}
\caption{Direct salol dielectric model comparison over the 21-observation equilibrium window. Bars show $\Delta$AIC relative to CPA+C; the dotted line marks $\Delta\mathrm{AIC}=10$.}
\label{fig:q1}
\end{figure}

A complementary dielectric-only comparison fits all four models directly to the 21-observation equilibrium record. CPA+C is selected by AIC and AICc, with $\Delta\mathrm{AIC}$ advantages of 27.5 over Avramov--Milchev, 35.0 over MYEGA, and 60.3 over VFT (Fig.~\ref{fig:q1}). This separate analysis tests whether CPA+C describes the broad dielectric trajectory without importing the viscosity-derived geometry.

\begin{figure}[H]
\centering
\includegraphics[width=.89\linewidth]{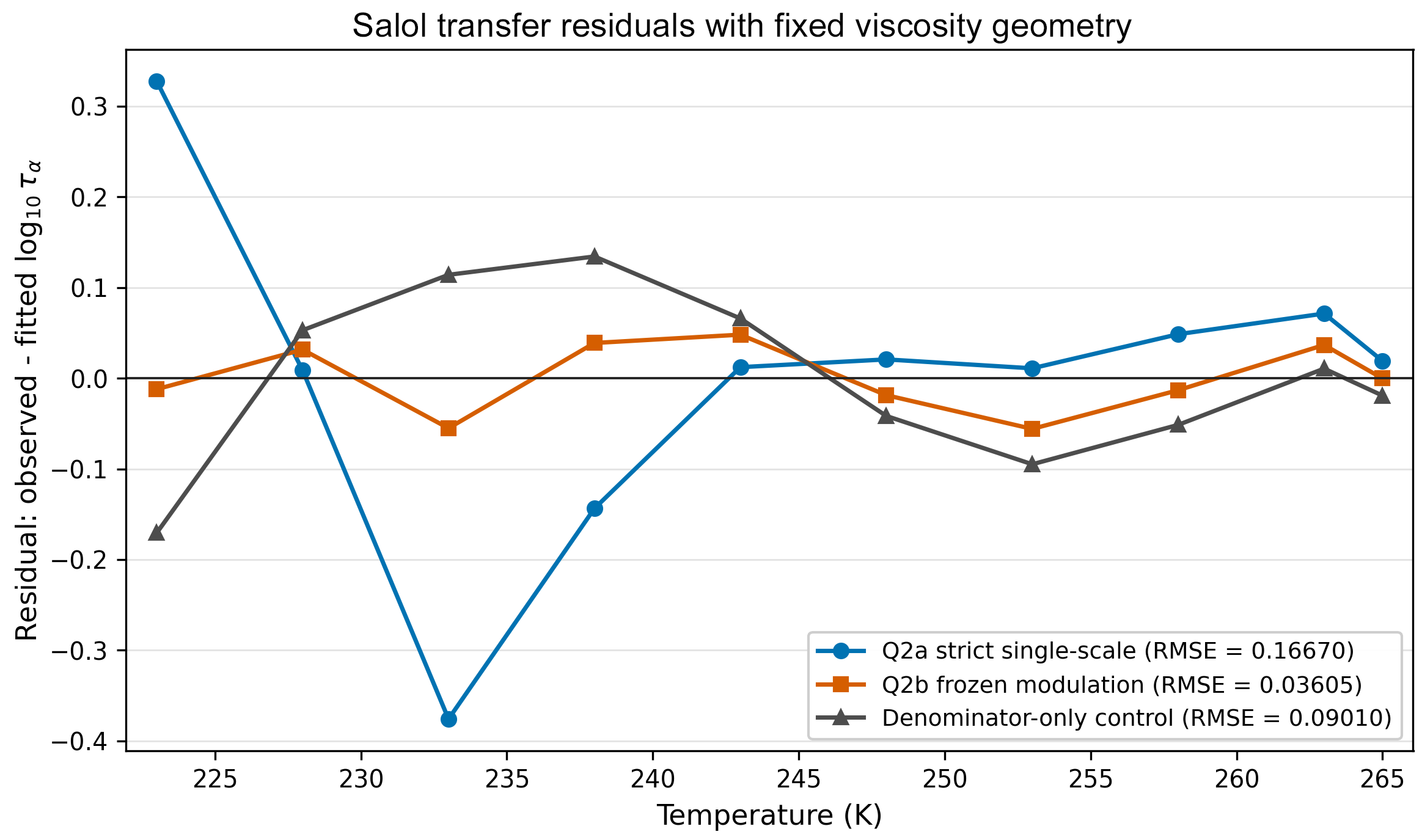}
\caption{Residuals in the salol transfer window comprising 10 dielectric observations. Q2b holds the viscosity-derived temperature geometry fixed and fits dielectric-specific coefficients; Q2a applies a single global offset and scale to the temperature-dependent part of the fitted viscosity law; CTRL retains the viscosity-fixed denominator while omitting $C_\eta(T)$.}
\label{fig:q2}
\end{figure}

The strict affine diagnostic (Q2a) assesses whether one global offset and scale applied to the temperature-dependent part of the fitted viscosity law is sufficient. At the point-estimate geometry, Q2a's AIC is 12.3040 units higher than CTRL's, and Q2a is favored over CTRL in 31 of the 1,000 propagated geometries. The supported transfer result is Q2b versus CTRL: the viscosity-derived temperature geometry remains fixed while Q2b fits its own coefficients to the dielectric relaxation data (Fig.~\ref{fig:q2}).

\section{Discussion}

\subsection{What the combined evidence establishes}

Three quantitative findings align. First, the 35-symbol salol viscosity record decisively favors CPA+C with the full five-parameter penalty retained. The preference persists in every symbol-level bootstrap resample and is accompanied by substantially lower held-out prediction error. Second, the source-verified B$_2$O$_3$ record favors the same rate-law construction in a chemically and structurally distinct glass former. Third, salol dielectric relaxation supports CPA+C at two levels: direct fitting favors it across the 21-observation equilibrium trajectory, and fixed-geometry transfer is favored over a control with the same denominator in the 10-observation window.

The fixed-geometry transfer is the more discriminating cross-observable test. Salol viscosity fixes $T_{\mathrm{lock},\eta}$, $T_{\mathrm{ref},\eta}$, and the full temperature dependence of $C_\eta(T)$ before the dielectric comparison, leaving only dielectric-specific coefficients to be estimated. Q2b's 16.32-unit AIC advantage over CTRL, together with preference for Q2b across all 1,000 propagated viscosity geometries, shows that the viscosity-derived modulation retains predictive information for molecular reorientation beyond the viscosity-fixed denominator alone.

The supported transfer is geometry-specific rather than a single global viscosity-to-dielectric scaling. Q2b preserves the viscosity-derived temperature geometry while allowing the dielectric observable its own coefficients. The weaker Q2a result shows that one global offset and scale are insufficient. The common empirical content lies in the fitted temperature dependence, while macroscopic flow and molecular reorientation respond to it with different amplitudes.

\subsection{Constraint modulation and CPA}

CPA treats the liquid’s physical organization as part of the conditions governing molecular rearrangement. Packing, bonding, local coordination, molecular arrangement, and cooperative motion help determine how readily the liquid can reconfigure. CPA+C tests whether the temperature dependence of that restriction appears in measured slowing.

The viscosity comparisons indicate why CPA+C obtains the better numerical performance. The salol and B$_2$O$_3$ records contain a change in curvature that the bounded modulation follows while the comparator equations leave more systematic residual structure. Above $T_{\mathrm{ref}}$, the constraint-coupled term changes with temperature; at $T_{\mathrm{ref}}$, its normalized weight saturates; continued slowing within $T_{\mathrm{lock}}<T\leq T_{\mathrm{ref}}$ then arises through the denominator. The empirical result concerns saturation of the fitted modulation, rather than saturation of the liquid's total physical constraint.

Salol and B$_2$O$_3$ select this construction with material-specific temperature coordinates and coupling amplitudes. Within salol, the viscosity-inferred $C_\eta(T)$ remains predictive for dielectric relaxation after dielectric-specific coefficients are fitted independently. The combined result is therefore a reproducible viscosity curvature across two materials together with a transferable fitted temperature dependence across two relaxation observables in salol.

\vspace{.8em}
\subsection{Operational coordinates and next tests}

$T_{\mathrm{lock}}$, $T_{\mathrm{ref}}$, and $C(T)$ are empirically inferred coordinates of the fitted rate law. Their stability under salol resampling and the successful transfer of the salol geometry make them concrete targets for comparison with independent measurements. The clearest physical test is to compare the inferred modulation with a structural or thermodynamic coordinate measured independently in the same material.

The CPA framing also suggests a broader experimental program. Because the current $C(T)$ depends only on temperature, controlled cooling-rate and aging protocols can test whether different paths to the same temperature alter the inferred geometry. Pressure, confinement, composition, and network connectivity offer independent ways to vary configurational access. Molecular simulation can compare the fitted modulation with changes in mobility and cooperative rearrangement. Fixed-geometry transfer in another glass former would provide a further cross-observable test when a sufficiently broad matched dataset becomes available.

\vspace{.8em}
\section{Conclusion}

As a glass-forming liquid cools, the rearrangements available to it become increasingly restricted. CPA+C gives the proposed rate consequence of that narrowing a measurable form. On source-verified salol and B$_2$O$_3$ viscosity records, the same bounded construction outperforms VFT, MYEGA, and Avramov--Milchev after full complexity penalties. In salol, it also predicts held-out viscosity observations more accurately and remains selected throughout the symbol-level resampling analysis.

Salol dielectric relaxation extends the result beyond viscosity. An independent dielectric fit favors CPA+C across the 21-observation equilibrium trajectory. More decisively, the salol viscosity fit fixes $T_{\mathrm{lock},\eta}$, $T_{\mathrm{ref},\eta}$, and the full temperature dependence of $C_\eta(T)$ before the dielectric coefficients are estimated. With that geometry held fixed, Q2b is favored over the denominator-matched control at the point estimate and across all 1,000 propagated viscosity geometries. The viscosity-derived temperature pattern therefore retains predictive information for molecular reorientation while the two observables retain their own fitted amplitudes.

These findings establish a reproducible constraint-modulated rate signature in the tested systems. CPA interprets that signature as a rate consequence of molecular reorganization under the liquid’s current physical constraints. Within the empirical law, $C(T)$ provides an operational bridge between that process hypothesis and measured slowing. Direct structural measurements, controlled variation of cooling history and constraint, and molecular simulation can now determine which physical quantity the modulation tracks.

\section*{Data and Reproducibility}

The data, code, canonical outputs, figures, validation records, and source-resolution documentation supporting this manuscript are archived in a consolidated Zenodo record~\cite{GavantPreckerCombinedGlassV5}. The record contains the frozen analysis specifications, primary inputs, fitting and verification code, canonical point-fit and symbol-bootstrap outputs, leave-one-symbol-out results, all 1,000 viscosity-to-dielectric transfer evaluations, the exact-source B$_2$O$_3$ audit, the OTP source-construction audit, figures, signed validation documents, and SHA-256 manifests.

The earlier viscosity and salol dielectric deposits remain available as historical records~\cite{GavantViscosityData,GavantDielectricData}. The manuscript source and compiled PDF were generated from the same finalized numerical outputs.

\section*{Acknowledgments}

No external grant funding was received for this study. 

OpenAI ChatGPT and Codex assisted with computational workflow development, numerical cross-checking, figure preparation, and manuscript editing. The authors reviewed all reported results against the governing source data, finalized computational records, and independent validation materials. The authors assume responsibility for the final analyses, interpretations, and text.

\section*{Author Contributions}
D.S.G.: Conceptualization, Methodology, Investigation, Formal analysis, Data curation, Visualization, Project administration; Writing: original draft, review and editing. 

C.E.P.: Validation, Investigation, Formal analysis, Data curation;
Writing: review and editing.

\appendix

\section{CPA+C Point-Fit Parameters}

Table~\ref{tab:params} lists the point parameters used for the principal viscosity figures. The salol fit occupies the $A=0$ boundary within numerical precision. The B$_2$O$_3$ solution is interior for both amplitude terms.

\begin{table}[!ht]
\centering
\caption{CPA+C point-fit parameters for the principal viscosity records.}
\label{tab:params}
\small
\begin{tabular}{lrrrrr}
\toprule
Record & $a$ & $A$ (K) & $B$ (K) & $T_{\mathrm{lock}}$ (K) & $T_{\mathrm{ref}}$ (K) \\
\midrule
Salol, 35 symbols & $-3.94554$ & $5.91\times10^{-29}$ & 582.3710 & 176.9716 & 234.0078 \\
B$_2$O$_3$, Table I & $-0.66575$ & 1295.8163 & 346.5765 & 424.7730 & 636.3627 \\
\bottomrule
\end{tabular}
\end{table}


\begin{thebibliography}{99}

\bibitem{Vogel1921}
H.~Vogel, ``Das Temperaturabh\"angigkeitsgesetz der Viskosit\"at von Fl\"ussigkeiten,'' \emph{Physikalische Zeitschrift} \textbf{22}, 645--646 (1921).

\bibitem{Fulcher1925}
G.~S.~Fulcher, ``Analysis of recent measurements of the viscosity of glasses,'' \emph{Journal of the American Ceramic Society} \textbf{8}, 339--355 (1925).

\bibitem{Tammann1926}
G.~Tammann and W.~Hesse, ``Die Abh\"angigkeit der Viscosit\"at von der Temperatur bei unterk\"uhlten Fl\"ussigkeiten,'' \emph{Zeitschrift f\"ur Anorganische und Allgemeine Chemie} \textbf{156}, 245--257 (1926).

\bibitem{Mauro2009}
J.~C.~Mauro, Y.~Yue, A.~J.~Ellison, P.~K.~Gupta, and D.~C.~Allan, ``Viscosity of glass-forming liquids,'' \emph{Proceedings of the National Academy of Sciences USA} \textbf{106}, 19780--19784 (2009), \url{https://www.pnas.org/doi/full/10.1073/pnas.0911705106}.

\bibitem{Avramov1988}
I.~Avramov and A.~Milchev, ``Effect of disorder on diffusion and viscosity in condensed systems,'' \emph{Journal of Non-Crystalline Solids} \textbf{104}, 253--260 (1988).

\bibitem{GavantDPTI}
D.~S.~Gavant, \emph{Dynamic Present Theory I: Foundations of Continuous Present Actualization}, version 7, Zenodo (2026), \url{https://doi.org/10.5281/zenodo.22649766}.

\bibitem{Casalini2016} R.~Casalini, S.~S.~Bair, and C.~M.~Roland, ``Density scaling and decoupling in $o$-terphenyl, salol, and dibutylphthalate,'' \emph{Journal of Chemical Physics} \textbf{145}, 064502 (2016), \url{https://doi.org/10.1063/1.4960513}. 

\bibitem{GavantDPTIv6Prospective}
D.~S.~Gavant, \emph{Dynamic Present Theory I: Foundations of Continuous Present Actualization}, version 6, Zenodo (2026), \url{https://doi.org/10.5281/zenodo.20798793}.

\bibitem{GavantGlassProtocol}
D.~S.~Gavant, \emph{Glass-Freeze Analysis Protocol v0}, Zenodo (2025), \url{https://doi.org/10.5281/zenodo.17502949}.

\bibitem{Adam1965}
G.~Adam and J.~H.~Gibbs, ``On the temperature dependence of cooperative relaxation properties in glass-forming liquids,'' \emph{Journal of Chemical Physics} \textbf{43}, 139--146 (1965).

\bibitem{Phillips1979}
J.~C.~Phillips, ``Topology of covalent non-crystalline solids I: Short-range order in chalcogenide alloys,'' \emph{Journal of Non-Crystalline Solids} \textbf{34}, 153--181 (1979).

\bibitem{Thorpe1983}
M.~F.~Thorpe, ``Continuous deformations in random networks,'' \emph{Journal of Non-Crystalline Solids} \textbf{57}, 355--370 (1983).

\bibitem{Gotze1992}
W.~G\"otze and L.~Sj\"ogren, ``Relaxation processes in supercooled liquids,'' \emph{Reports on Progress in Physics} \textbf{55}, 241--376 (1992).

\bibitem{Laughlin1972}
W.~T.~Laughlin and D.~R.~Uhlmann, ``Viscous flow in simple organic liquids,'' \emph{Journal of Physical Chemistry} \textbf{76}, 2317--2325 (1972), \url{https://doi.org/10.1021/j100660a023}.

\bibitem{Napolitano1965}
A.~Napolitano, P.~B.~Macedo, and E.~G.~Hawkins, ``Viscosity and Density of Boron Trioxide,'' \emph{Journal of the American Ceramic Society} \textbf{48}, 613--616 (1965), \url{https://doi.org/10.1111/j.1151-2916.1965.tb14690.x}.

\bibitem{Burnham2002}
K.~P.~Burnham and D.~R.~Anderson, \emph{Model Selection and Multimodel Inference}, 2nd ed. (Springer, New York, 2002), \url{http://dx.doi.org/10.1007/b97636}.

\bibitem{Hansen1998}
C.~Hansen, F.~Stickel, R.~Richert, and E.~W.~Fischer,
``Dynamics of glass-forming liquids. IV. True activated behavior above 2 GHz in the dielectric $\alpha$-relaxation of organic liquids,''
\emph{Journal of Chemical Physics} \textbf{108}, 6408--6415 (1998),
\url{https://doi.org/10.1063/1.476063}.

\bibitem{Lunkenheimer2010} P.~Lunkenheimer, S.~Kastner, M.~K{\"o}hler, and A.~Loidl, ``Temperature development of glassy $\alpha$-relaxation dynamics determined by broadband dielectric spectroscopy,'' \emph{Physical Review E} \textbf{81}, 051504 (2010), \url{https://doi.org/10.1103/PhysRevE.81.051504}.

\bibitem{GavantPreckerCombinedGlassV5}
D.~S.~Gavant and C.~E.~Precker,
``A Constraint-Modulated Rate Law for Glass-Forming Systems: Viscosity Evidence and Dielectric Transfer in Salol,'' version~5, Zenodo (2026),
\url{https://doi.org/10.5281/zenodo.22306850}.

\bibitem{GavantViscosityData}
D.~S.~Gavant and C.~E.~Precker, ``A Constraint-Modulated Viscosity Law for Broad-Window Glass-Forming Systems: data and code,'' Zenodo (2026), \url{https://doi.org/10.5281/zenodo.20572383}.

\bibitem{GavantDielectricData}
D.~S.~Gavant, ``CPA+C Viscosity-to-Dielectric Transfer in Salol: data, code, and validation record,'' Zenodo (2026), \url{https://doi.org/10.5281/zenodo.21831309}.

\end{thebibliography}
\end{document}